\documentclass{pasj00} 
\usepackage{mathrsfs}
\begin{document} 

\SetRunningHead{Yoshiaki {\sc Sofue}}{Dark Halo of M31 and the Galaxy}
\Received{} \Accepted{} 

\def\kms{km s$^{-1}$} \def\Msun{M_\odot}
\def\be{\begin{equation}} \def\ee{\end{equation}}
\def\bc{\begin{center}} \def\ec{\end{center}} 
\def\ab{a_{\rm b}} \def\ad{a_{\rm d}} 
\def\Mb{M_{\rm b}} \def\Md{M_{\rm d}} 
\def\Mh{M_{\rm h}} 
\def\Mbd{M_{\rm b+d}}  \def\dV{de Vaucouleurs } \def\dv{de Vaucouleurs} 
\def\ha{H$\alpha$}

\title{Dark Halos of M31 and the Milky Way}
\author{Yoshiaki {\sc Sofue} } 
\affil{ Institute of Astronomy, University of Tokyo, Mitaka, 181-0015 Tokyo \\
Email:{\it sofue@ioa.s.u-tokyo.ac.jp}
 }

\maketitle

\begin{abstract} 

Grand rotation curves (GRC) within $\sim 400$ kpc of M31 and the Milky Way were constructed by combining disk rotation velocities and radial velocities of satellite galaxies and globular clusters. The GRC for the Milky Way was revised using the most recent Solar rotation velocity. The derived GRCs were deconvolved into a de Vaucouleurs bulge, exponential disk, and a dark halo with the Navarro-Frenk-White (NFW) density profile by the least $\chi^2$ fitting. Comparison of the best-fit parameters revealed similarity of the disks and bulges of the two galaxies, whereas the dark halo mass of M31 was found to be twice the Galaxy's.  We show that the NFW model may be a realistic approximation to the observed dark halos in these two giant spirals.

\end{abstract}

\section{Introduction}
Rotation curve of the Andromeda galaxy M31 (NGC 224) has been obtained in details and are used for determining the mass distribution in the disk and dark halo (Sofue et al. 1999; Carignan et al. 2006; Chemin et al. 2009; Corbelli et al. 2011). The dark halo mass beyond the disk has been derived using kinematics of satellite galaxies (e.g., Metz et al. 2007; van der Marel et al. 2008; Tollerud et al. 2012). Globular clusters are also used extensively for the mass determination of dark halo (Veljanovski et al. 2010) as well as for inner kinematics (Galletti et al. 2004).

In our recent work on the Milky Way Galaxy, we constructed a running averaged rotation curve for a wide region from the Galactic Center to outermost dark halo, which we called the grand (or pseudo) rotation curve (hereafter, GRC: Sofue 2012, 2013). In the present paper, we revise the GRC by adopting the most recent value of the Solar rotation velocity, $V_0=238$ \kms, from VERA observations (Honma et al. 2012).

We apply a common method to construct GRCs to M31 and the Milky Way by combining disk rotation curves and radial velocities of satellite galaxies and outer globular clusters. We deconvolve the GRCs into bulge, disk and dark halo, and determine the dynamical parameters. We compare the results for M31 and the Milky Way, and discuss the similarity of the two galaxies. It is emphasized that the NFW model indeed works in largely extended dark halos of real galaxies.

\section{Grand Rotation Curves}

\subsection{M31}

The rotation curve of the disk of M31 has been obtained by many authors using HI, CO and \ha \ line kinematical data as listed in table 1. For the disk region, we simply adopt the averaged rotation velocity $V(R)$ as a function of radius $R$ from these data. For the halo region beyond the disk, we employ kinematics of non-coplanar objects orbiting M31, which include satellite galaxies and globular clusters (table 1). We adopt a distance to M31 of 770 kpc, and the systemic velocity of 300 \kms (Courteau and van den Bergh 1999)

\begin{table*} 
\bc
\caption{References to the data in figure \ref{runall} }
\begin{tabular}{lll}
\hline\hline  
Rectangles at $R<10$ kpc & Disk RC (CO) & Loinard et al. (1995) \\
Grey circles at $R<32$ kpc & ibid (combi: HI, CO, opt) & Sofue et al. (1981, 1999) \\
Grey open circles linked by line & ibid (HI) & Carignan et al. (2006)\\
Black-grey circles linked by line & ibid  (HI, CO) & Chemin et al. (2009) \\
Grey reverse triangles linked by line & ibid (HI) & Corbelli et al. (2010)\\
\hline
Rectangles at $R>40$ kpc & galaxies around M31 & Metz et al. (2007) \\
Triangles with bars & ibid  &van der Marel et al. (2008) \\
Reverse triangle at $R>40$ kpc & ibid  & Tollerud et al. (2012) \\
Open circles with bars & Globular clusters & Veljanovski et al. (2014) \\
\hline
\end{tabular}
\ec
\label{reference} 
\end{table*}

We assume that the distribution of non-coplanar objects is spherical around M31 and the velocity vectors are random. We define a pseudo rotation velocity, $V$, as the velocity that yields the enclosed mass $M(R)$ within a radius $R$ from the galaxy's center by 
\be
M(R)=V^2 R/G,
\ee
where $G$ is the gravitational constant.  The pseudo rotation velocity is here replaced by the Virial velocity for an ensamble of particles orbiting around the center of mass in random orbits (Limber and Mathews 1960; Bahcall and Tremaine 1981). If the orbits are random, the velocity is evaluated by 
\be
V=\sqrt{3<v_z^2>},
\label{pse}
\ee
where $v_z$ is line-of-sight velocity of each object (Limber and Mathews 1960), and $<v_z^2>$ is the mean of squares of $v_z$ in the ensamble. The factor $\sqrt{3}$ was multiplied for correction for the degree of freedom of random motion (Limber and Mathews 1960). 

If the orbits are not random, the above estimation may be corrected for the orbits shapes by 
\be
V=C\sqrt{3<v_z^2>}.
\label{pse}
\ee
Here, the coefficient $C$ is a factor ranging from $\sqrt{3}\pi/8\sim 0.7$ for radial orbits to $3\pi/8\sim 1.2$ for circular orbits (Bahcall and Tremaine 1981), depending on the shapes of orbits as well as on the assumed potential and distribution function of objects (Evans et al. 2003). 

Since, it is rather the potential that we want to determine in this paper, and the other factors are unknown, we here assume that $C=1$. It should be mentioned that the obtained result in this paper depends on this correction factor in the sense that the dark halo mass is proporional to $C^2$. Namely, the mass could be changed by a factor of $\sim 2$. Hence, the given errors in this paper are statistical values, not considering the systematic effects by the assumption.

For objects whose three dimensional coordinates are unknown, the distance from M31's center $R$ is related to the projected distance $r_{\rm proj}$ by
\be
R= {\pi \over 2}~ r_{\rm proj}.
\ee  
We adopt thus calculated pseudo rotation velocity as the rotation velocity.
 
If there exists systematic rotation, as observed for the globular clusters around M31 (Veljanovski et al 2014), a correction is necessary as  described in Appendix. We assume that the rotation axis is parallel to the disk's rotation axis, $i=77^\circ$, so that the correction is small and equation \ref{pse} holds also for the globular cluster system.
 
\begin{figure}
\begin{center} 
(a)\includegraphics[width=8cm]{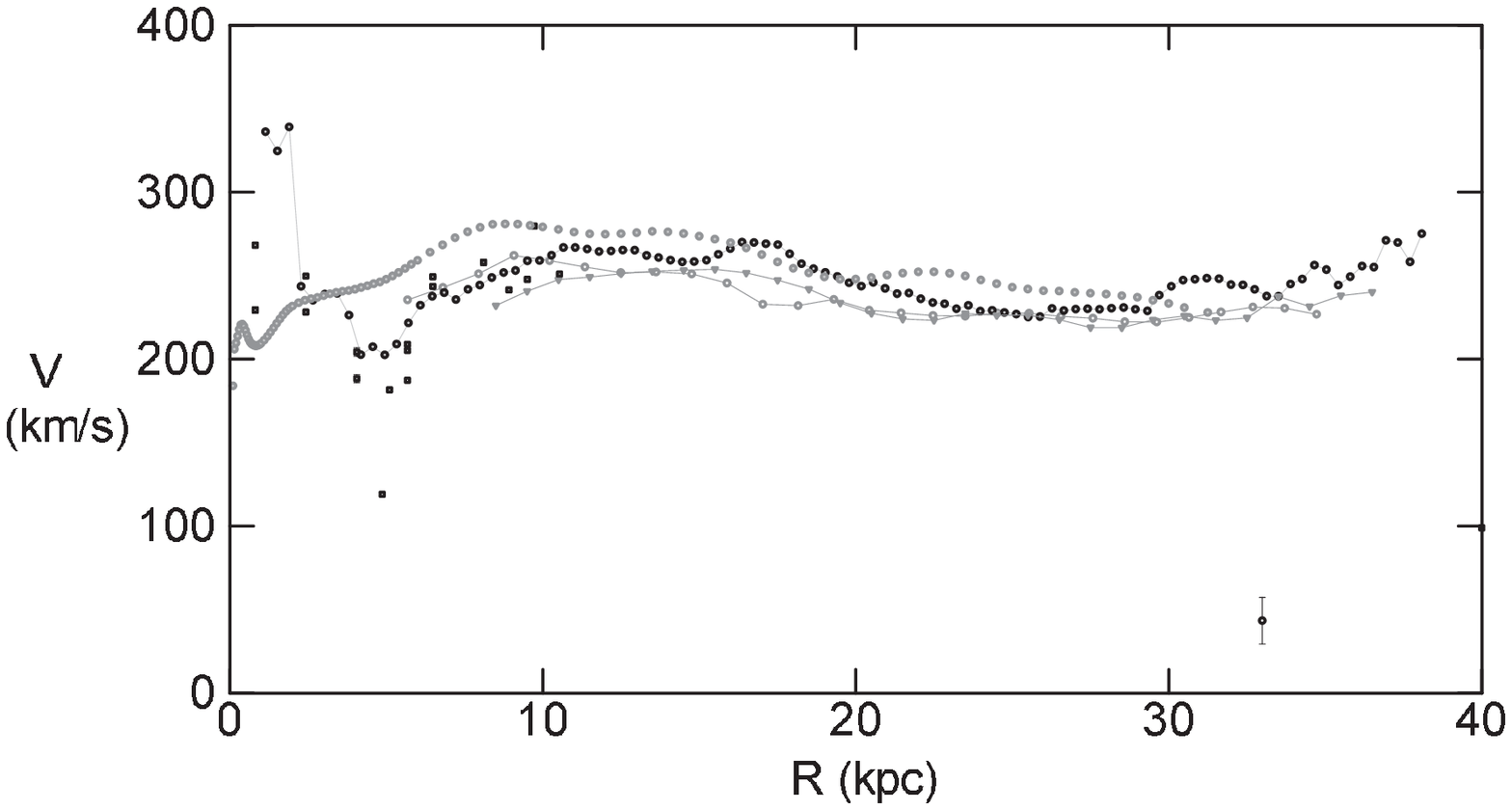}  
(b)\includegraphics[width=8cm]{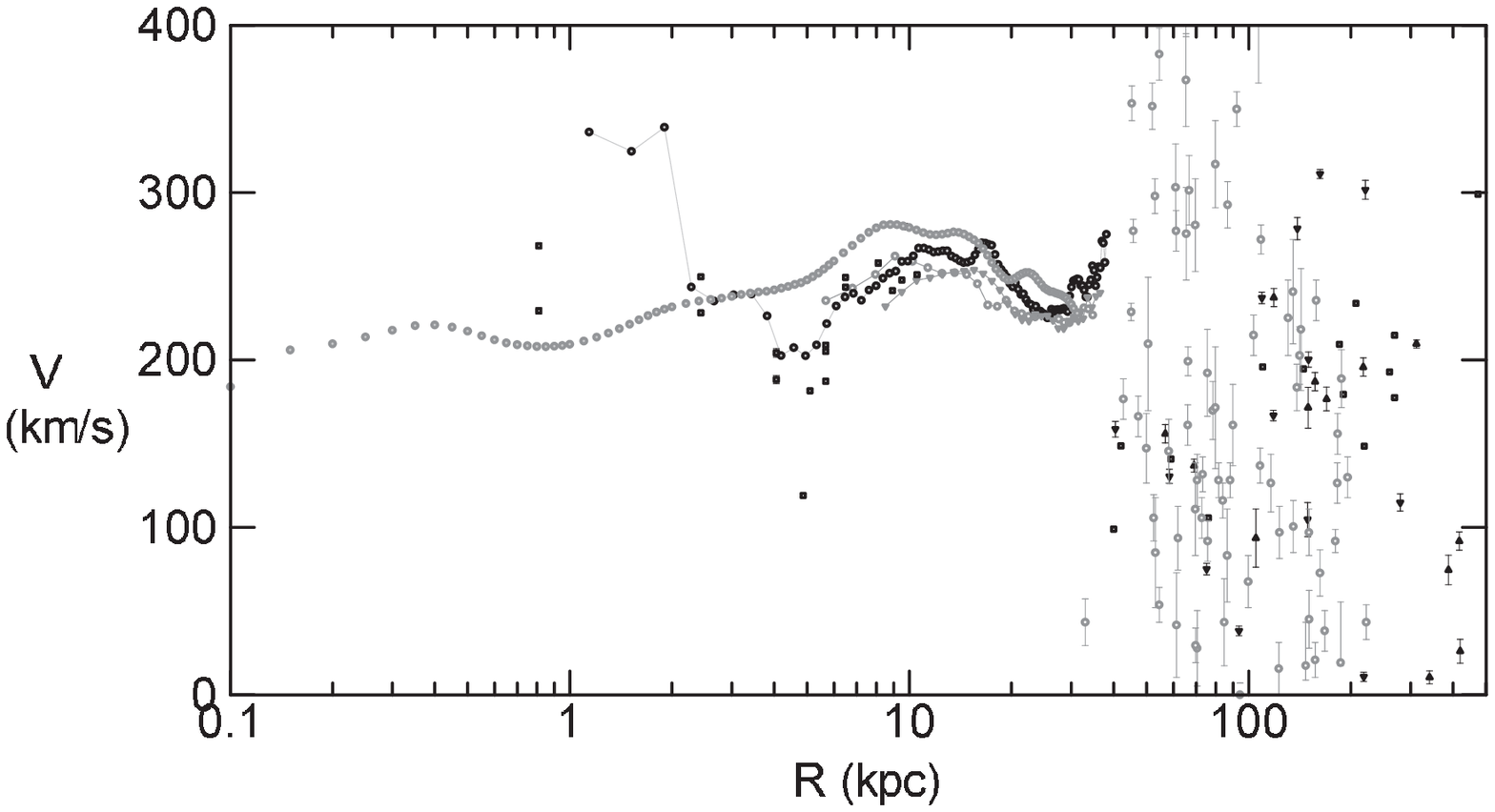}  
\end{center}  
\caption{Rotation velocities of the disk and pseudo rotation velocities of non-coplanar objects in M31 (a) in linear and (b) semi-logarithmic scaling. References to the data are listed in table 1.}
\label{runall} 
\end{figure}

Figure \ref{runall} shows compiled rotation velocities for the disk, and pseudo rotation velocities of satellites and globular clusters within radius $500$ kpc from M31. Using these velocities, we calculate running averaged velocities in each bin at logarithmic interval between $R$ and $1.2 R$ with $R$ starting from $R=0.1$ kpc by applying the Gaussian running averaging procedure described in Sofue (2012, 2013). 
Note that we here use the root of the mean, $\sqrt{<v_z^2>}$ as the pseudo rotation velocity, because the squared velocity is more directly related to the Virial mass according to equation \ref{pse}, whereas we took $<v_z>$ in the ealier work.

\begin{figure}
\begin{center}   
(a)\includegraphics[width=8cm]{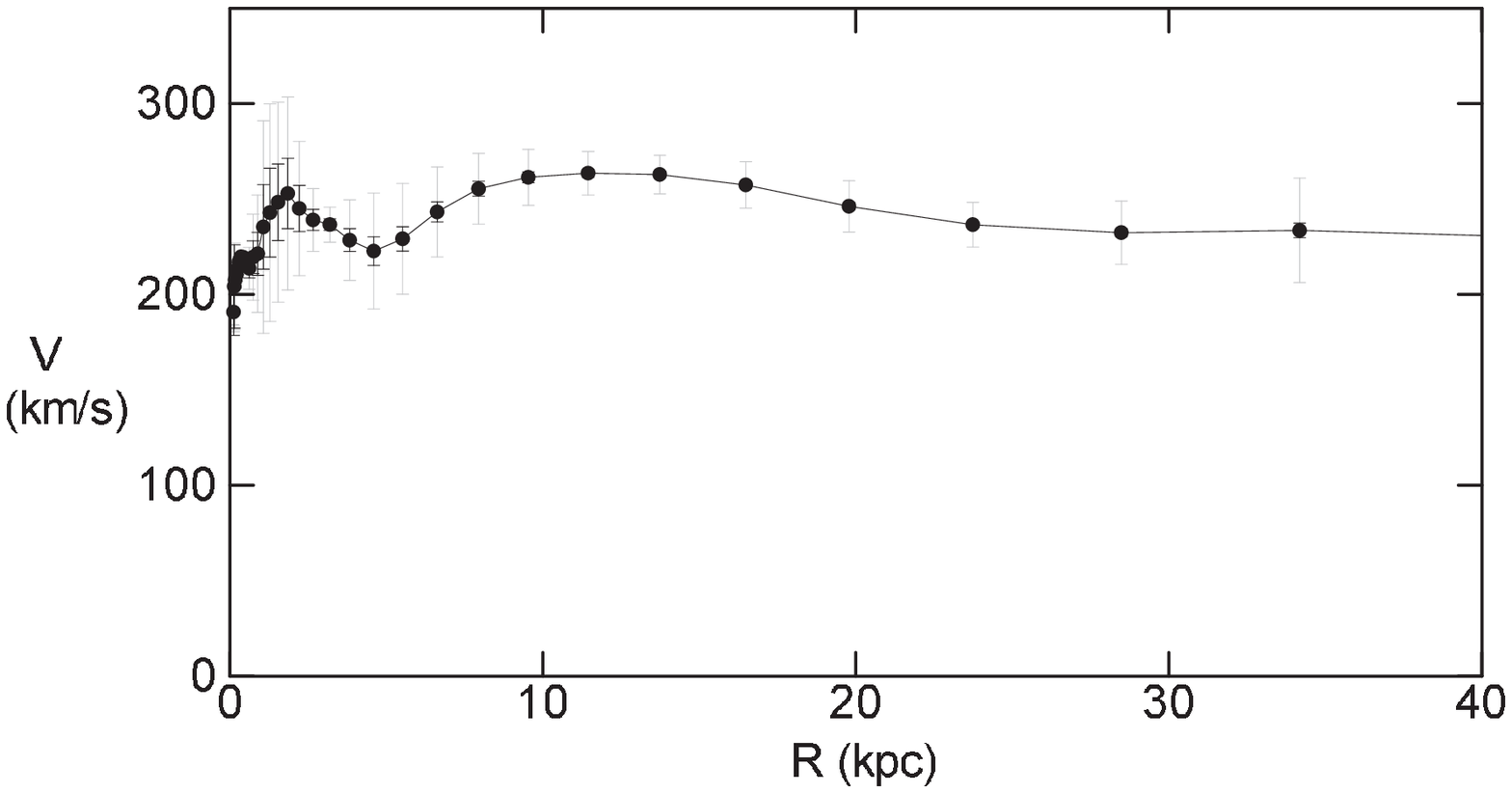}  
(b)\includegraphics[width=8cm]{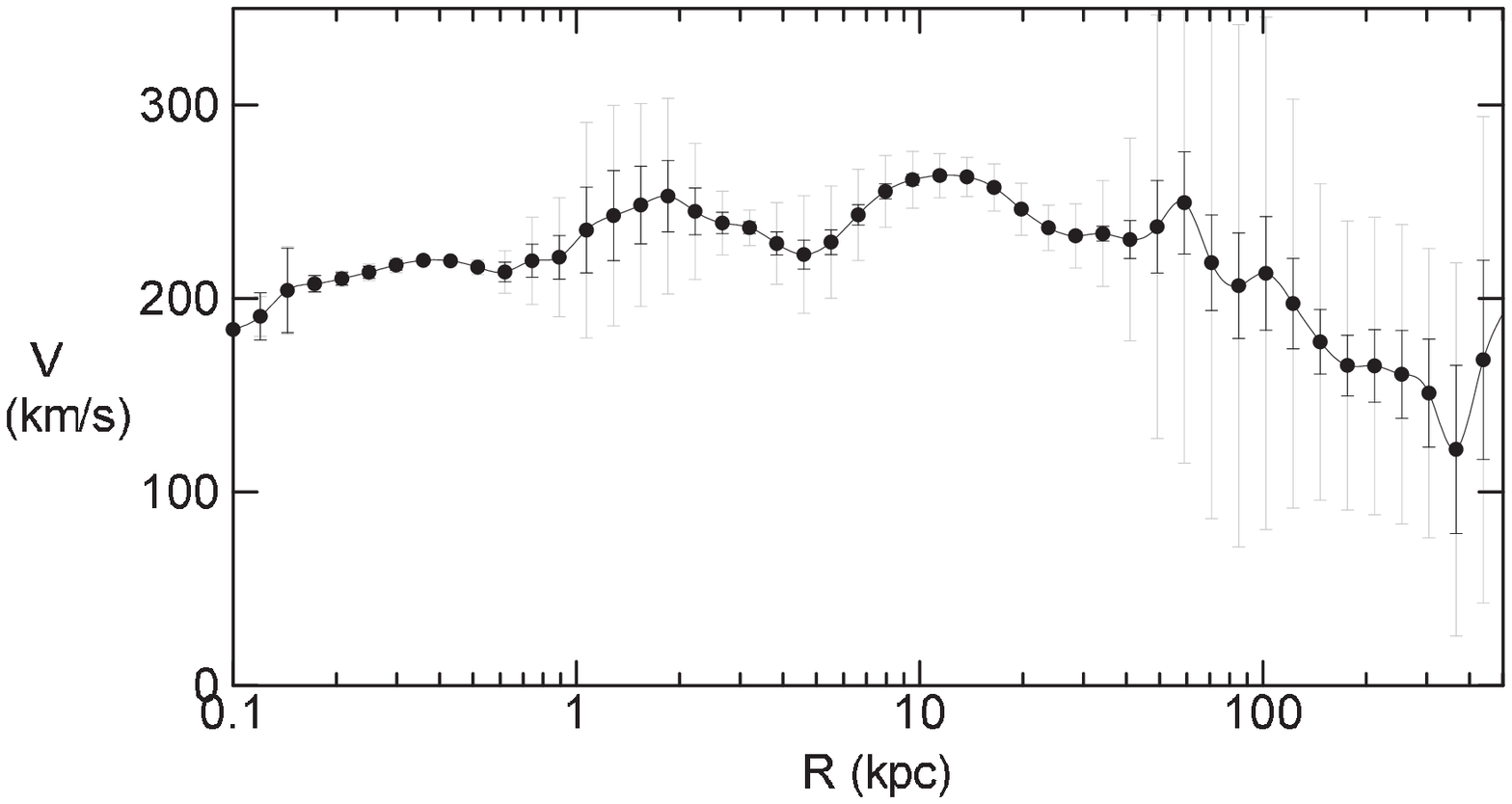}
\end{center}  
\caption{GRC of M31, showing running averaged values of pseudo rotation velocities in (a) linear and (b) logarithmic scalings. Long and short bars represent modified standard deviations and errors, respectively. } 
\label{rc385} 

\begin{center} 
(a)\includegraphics[width=8cm]{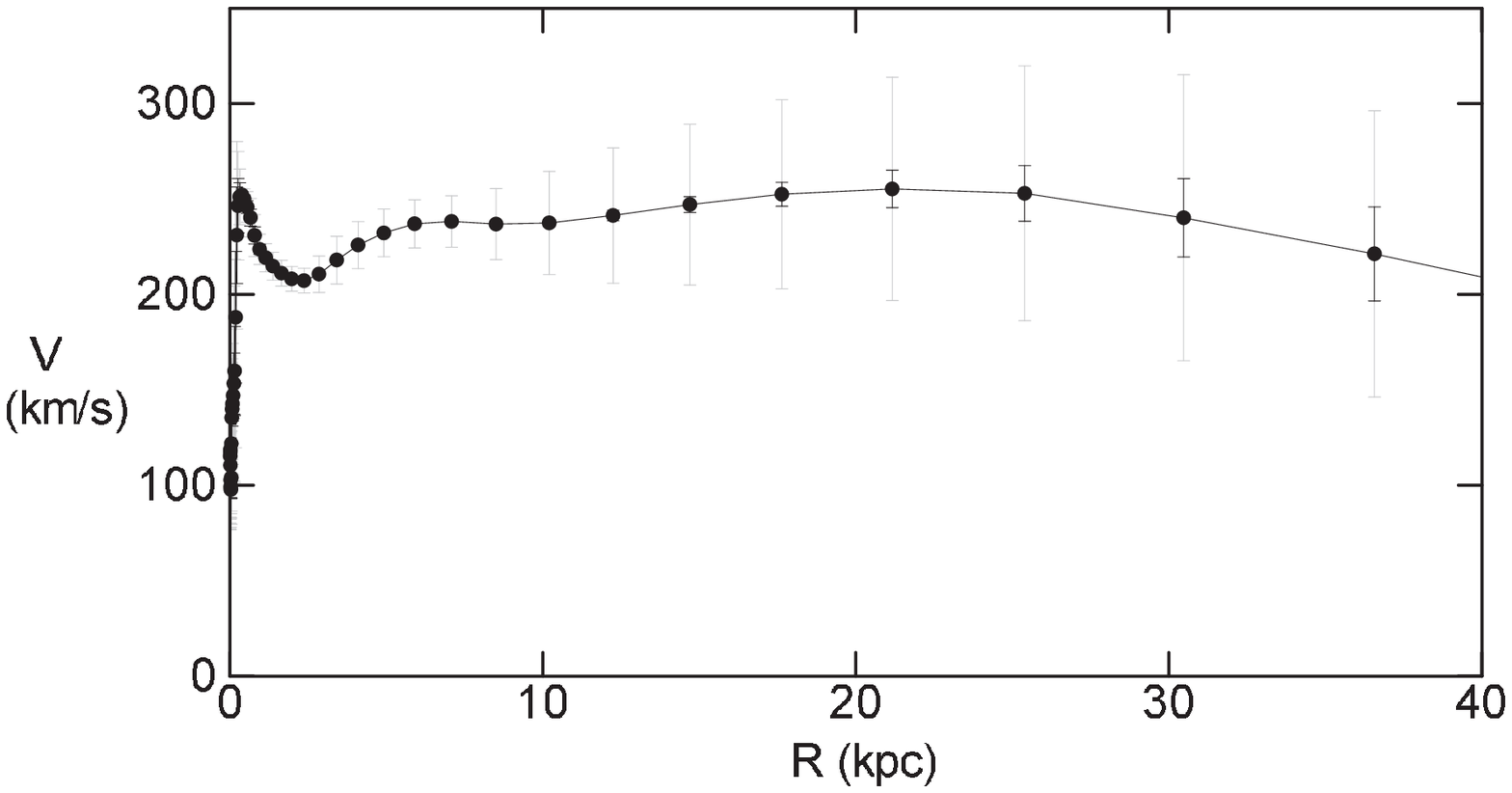} 
(b)\includegraphics[width=8cm]{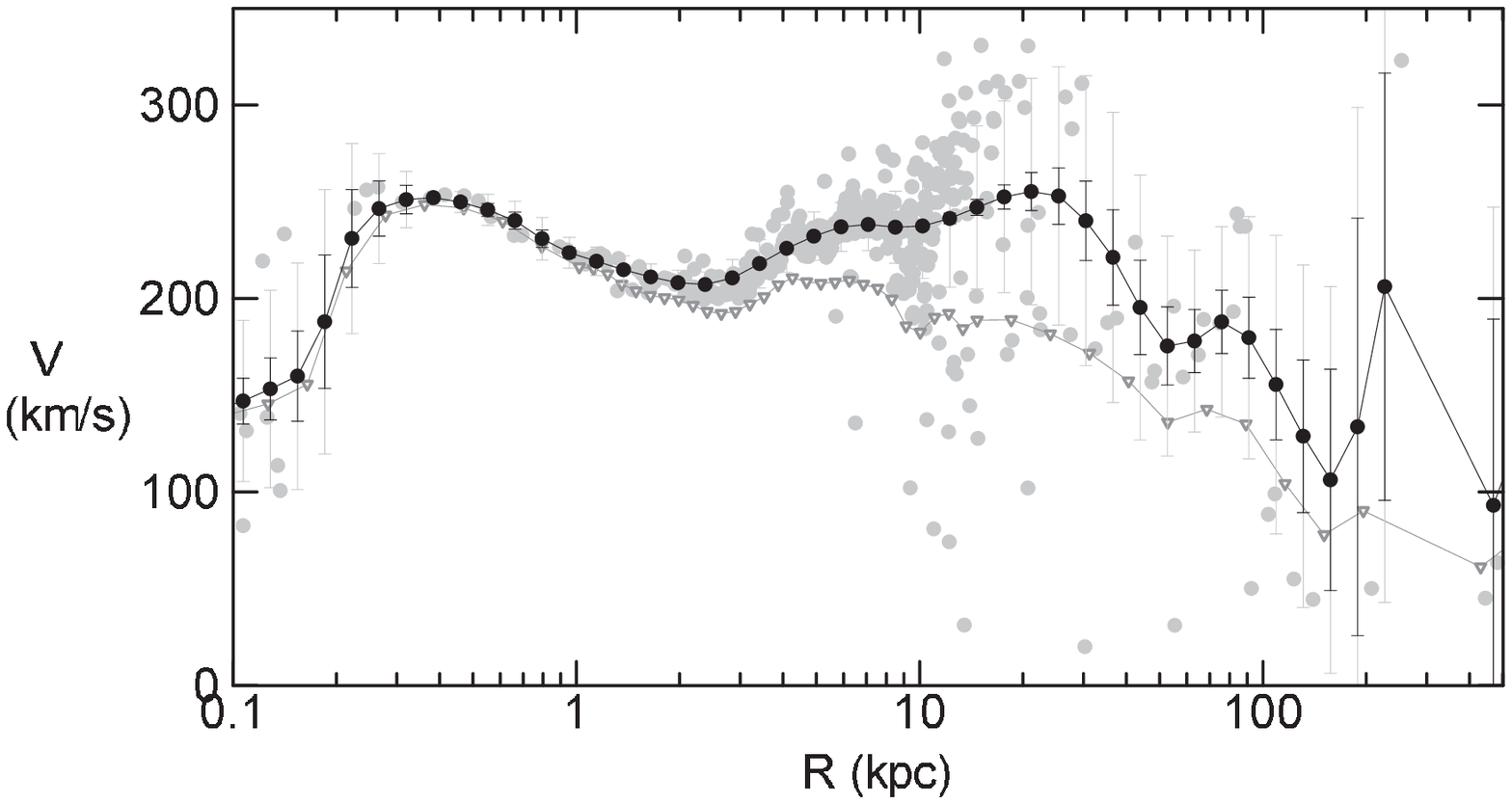}  
\end{center}
\caption{New GRC of the Milky Way in (a) linear and (b) semi-logarithmic scalings. Grey dots are the data for individual objects, and grey line with triangles shows our earlier GRC (Sofue 2012, 2013).} 
\label{rc385mw} 
\end{figure}

The pseudo rotation velocities are plotted in figure \ref{rc385}, where long bars represent modified standard deviations among the used data in each bin, and short bars are modified standard errors. 
Here, the modified deviation and error are defined by $s_d=\delta V \simeq \delta V^2/2V = s_{d:V^2}/2V$ and $s_e=s_{e:V^2}/2V$, recalling the propagation of the deviation (error) of $V^2$ and $V$ by the derivative relation $\delta V^2=2V \delta V$. Here, $s_{d:V^2} = (< (3v_z^2 - V^2)^2>)^{1/2}$ and $s_{e:V^2}=s_{d:V^2}/N^{1/2}$ are the standard deviation and error around $V^2$ in each radius bin, respectively, and $N$ is the number of data points in the bin. 

These definitions were employed for the reason why we chose the square value, $v_z^2$, as the independent variable, instead of the linear value, $v_z$. We call thus obtained plot of $V$ the grand rotation curve (GRC). 

\begin{figure}
\begin{center}  
(a)\includegraphics[width=8cm]{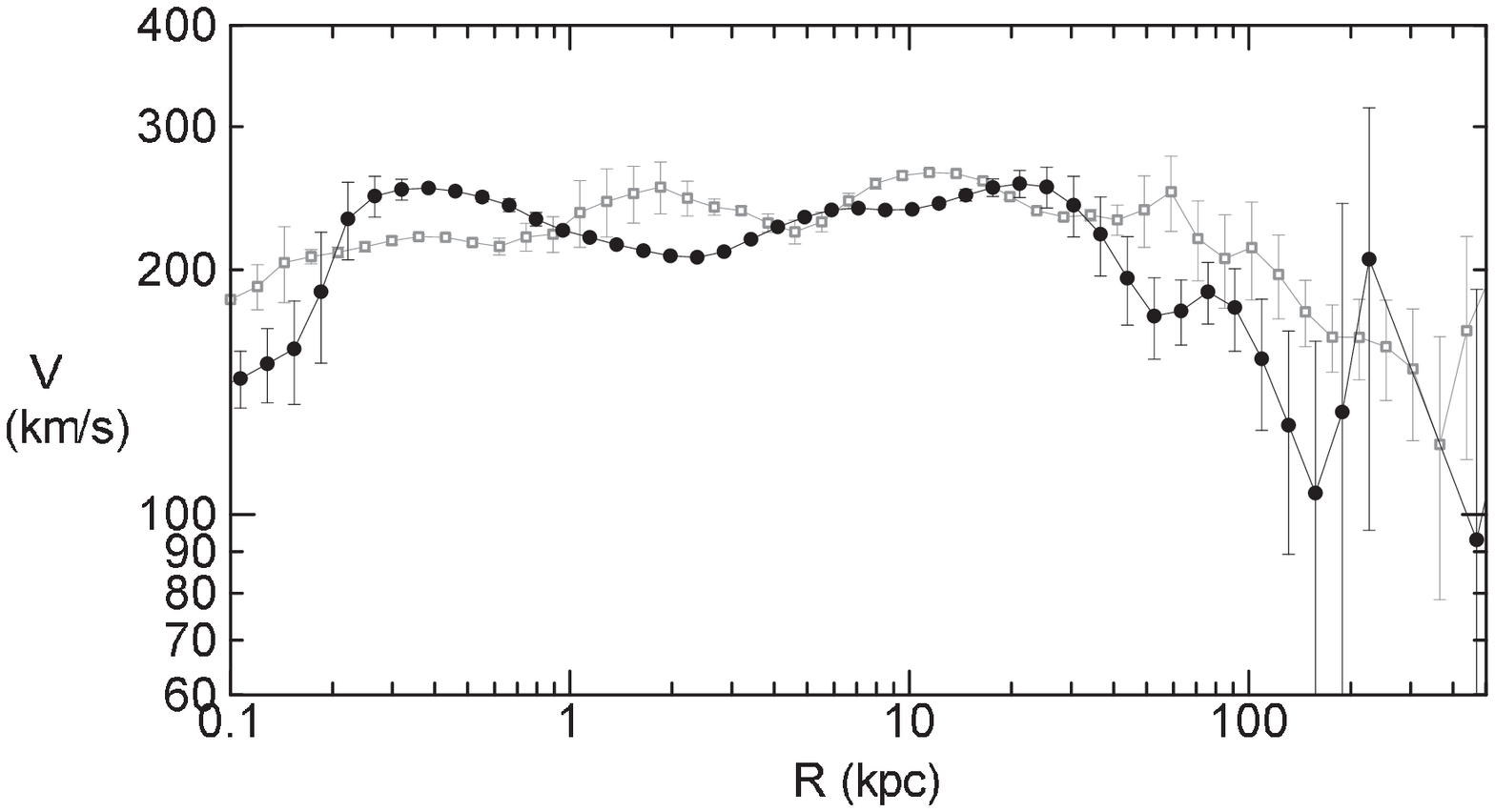}  
(b)\includegraphics[width=8cm]{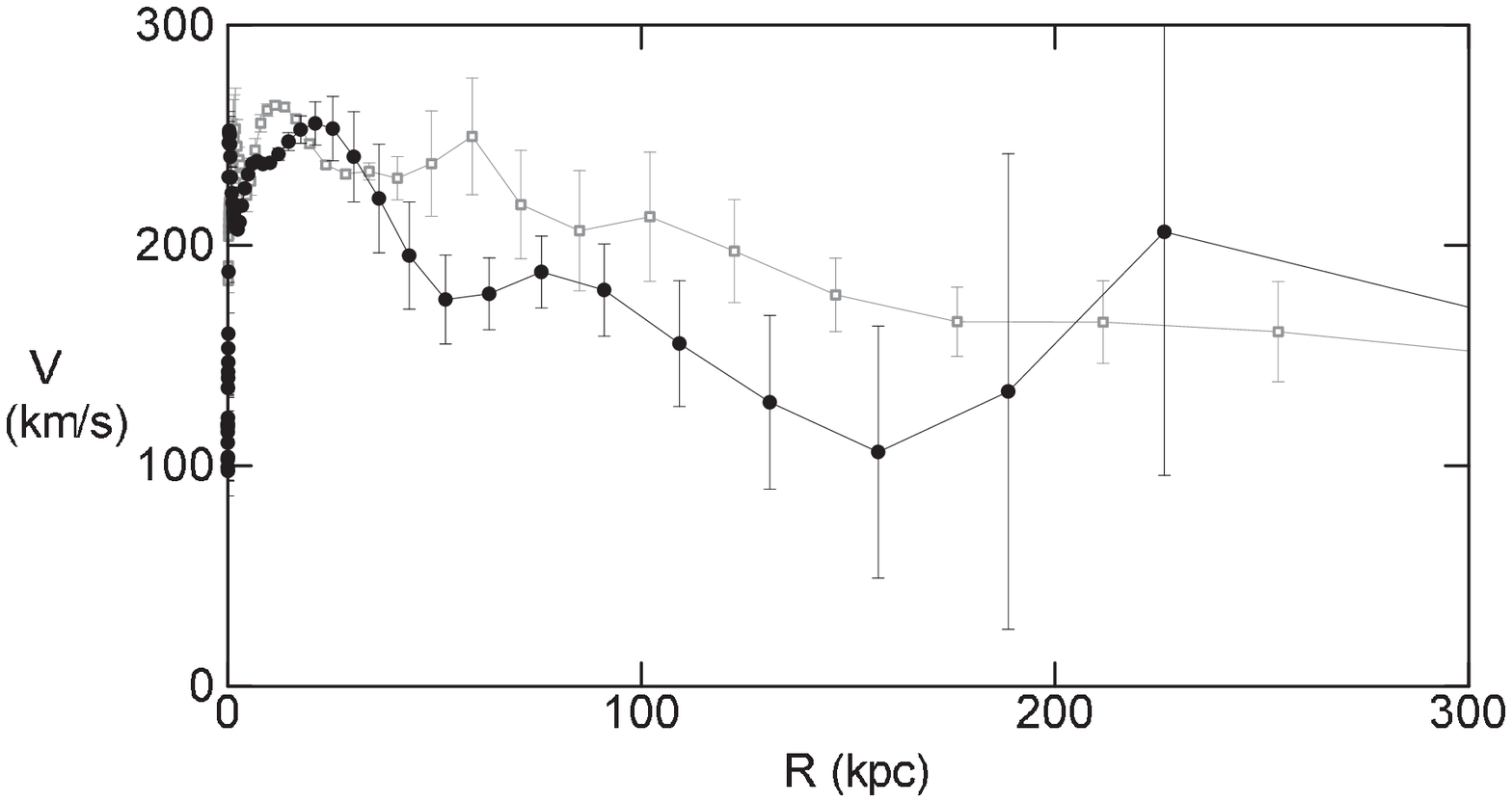}  
\end{center}
\caption{New GRC of the Milky Way (full line) compared with that of M31 (grey line) in (a) logarithmic and (b) linear scalings. The bars are modified standard errors.} 
\label{M31vsMW} 
\end{figure} 

\subsection{The Milky Way}

We revise the GRC of the Milky Way obtained in our earlier work (Sofue 2013) by adopting the most recently determined value of the Solar rotation velocity, $V_0=238$ \kms at  $R_0=8$ kpc, from VERA observations (Honma et al. 2012) in place of 200 \kms in the earlier paper. We also adopt the same correction factor for velocity as for M31, $V=\sqrt{3<v_z^2>}$, in place of $\sqrt{2} <v_z>$. In the earlier work, the degree of freedom of motion was assumed to be two, considering that each particle has transverse velocity same as the radial velocity. 

The revised GRC is shown in figure \ref{rc385mw}, where used data are superposed by grey dots, among which a few objects with $v_z \sim 400$ \kms have been removed from the analysis. The new GRC exhibits higher velocities than our eariler result shown by the grey line with triangles. The higher disk velocities are due to the adopted larger value of $V_0=238$ \kms as well as to higher velocities for non-coplanar objects because of here adopted correction of $\sqrt{3}$ instead of $\sqrt{2}$. Also, the presently employed average of the squared velocity $v_z^2$ lead to slightly higher mean velocity compared to the linear mean of $v_z$ in the previous work.

In figure \ref{M31vsMW} we compare the new GRC of the Milky Way with M31. The two GRCs show a remarkable similarity in the halo regions, indicating similarity dark matter distributions.

\subsection{Characteristics of the GRC}
The general characteristics of the obtained GRCs for M31 and the Galaxy may be summarized as follows:

 (i) Rotation curves in the bulges look typical in the plot up to 40 kpc, but the enlarged curves in logarithmic presentation exhibits that they have inner structures. Particularly, the innermost curve of the Milky Way was shown to be composed of multiple bulges with steep concentrations than the de Vaucouleurs law (Sofue 2013).

 (ii) The disk rotation curves are nearly flat, showing broad maxima at $R\sim 10$ to 20 kpc in both galaxies.

(iii) Beyond $R\sim 30$ kpc the rotation velocity declines smoothly until the edge of the dark halo. The slope is, however, milder than the Keplerian law, implying  that the halo cannot be represented by the Plummer or exponential type potentials that require steeper mass concentration. It will be shown in the next section that the NFW model is a good approximation to represent the observed GRCs in the halos.

\section{Deconvolution of GRC into Bulge, Disk and Dark Halo} 

We assume that the GRCs shown in figures \ref{rc385} and \ref{rc385mw} are composed of bulge, disk, and dark halo contributions as
\be
V(R)^2=V_{\rm b}(R)^2+V_{\rm d}(R)^2+V_{\rm h}(R)^2,
\label{vrot}
\ee 
where $V(R), ~V_{\rm b}(R),~ V_{\rm d}(R)$, and $V_{\rm h}(R)$ are the rotation velocity at galacto-centric distance $R$, and those for the bulge, disk and dark halo, respectively. By fitting the GRC with a model rotation curve using the the least $\chi^2$ method, following the method presented by  Sofue (2012, 2013), we searchd for the best-fit parameters of the bulge, disk and dark halo.
 
\subsection{Bulge}  
 
The bulge is assumed to have the \dV (1958) profile for the surface mass density as 
\be 
\Sigma_{\rm b}(r) = \Sigma_{\rm be} {\rm exp} 
\left[-\kappa \left\{\left(r / a_{\rm b} \right)^{1/4}-1\right\}\right],
\ee 
where $\kappa=7.6695$, $\Sigma_{\rm be}$ is the surface mass density at the half-mass scale radius $R=\ab$.  The total mass is calculated by 
\be
\Mb= 2 \pi \int_0^\infty r \Sigma_{\rm b}(r) dr 
=\eta \ab^2 \Sigma_{\rm be},
\ee
with $\eta=22.665$ being a dimensionless constant. 
The circular rotation velocity is then given by 
\be V_{\rm b}(R) = \sqrt{ G\Mb(R) / R}. 
\label{Vb}
\ee  

In the fitting procedure, $\Mb$ and $\ab$ are taken as the two free parameters. The bulge of our Galaxy was shown to be composed of multiple bulges with exponential density profiles, whereas the de Vaucouleurs law rather fails to reproduce the innermost rotation curve (Sofue 2013). Hence, the present analysis will be not accurate enough for the discussion of the bulge in the Milky Way. 

\subsection{Disk}

The galactic disk is approximated by an exponential disk, whose surface mass density is expressed as
\be 
\Sigma_{\rm d} (R)=\Sigma_0 {\rm exp} \left(  -{R /  \ad } \right), 
\label{disksigma}
\ee 
where $\Sigma_0$ is the central value and  $\ad$ is the scale radius. The total mass of the exponential disk is given by 
\be
\Md= \int_0^\infty 2 \pi r \Sigma_{\rm d} dr=2 \pi \Sigma_0 \ad^2.
\ee
The rotation curve for a thin exponential disk is expressed by
\be
V_{\rm d} (R)=\sqrt{ { G \Md /  \ad}} {\mathscr{D}}(X),
\label{Vd}
\ee
where  $X=R/\ad$, and $ \mathscr{D}(X)$ is the expression obtained by Freeman (1970) for a flat exponential disk. As the two free parameters we chose $\Md$ and $\ad$.

\subsection{Dark Halo}

For the dark halo, three mass models have been so far proposed: the semi-isothermal (Begeman et al. 1991),  NFW (Navarro, Frenk and White 1996), and Burkert (1996) models. The outermost rotation curves in figures \ref{fit385} and \ref{fit385mw} are not flat at all, so that the isothermal model is not a good approximation. Since the NFW and Burkert models are essentially the same except for the very central part, we here adopt the NFW profile. The NFW density profile is expressed as 
\be 
\rho(R)={\rho_0 /[  X\left(1+ X \right)^2] } ,
\label{nfw}
\ee 
where $X={R/ h}$, and $\rho_0$ and $h$ are the representative (scale) density and scale radius of the dark halo, respectively. In the fitting procedure, we chose $\rho_0$ and $h$ as the two free parameters.

The enclosed mass within radius $R$ is given by
\be 
\Mh (R)  
 = 4 \pi \rho_0 h^3 \left\{ {\rm ln} (1+X)-{X /( 1+X)}\right\}.
\label{mh} 
\ee
The circular rotation velocity is given by
\be
V_{\rm h} (R)=\sqrt{G \Mh (R) / R} .
\label{vh}
\ee

\subsection{Fitting result by the least $\chi^2$ method}

\begin{figure}
\begin{center}    
(a)\includegraphics[width=8cm]{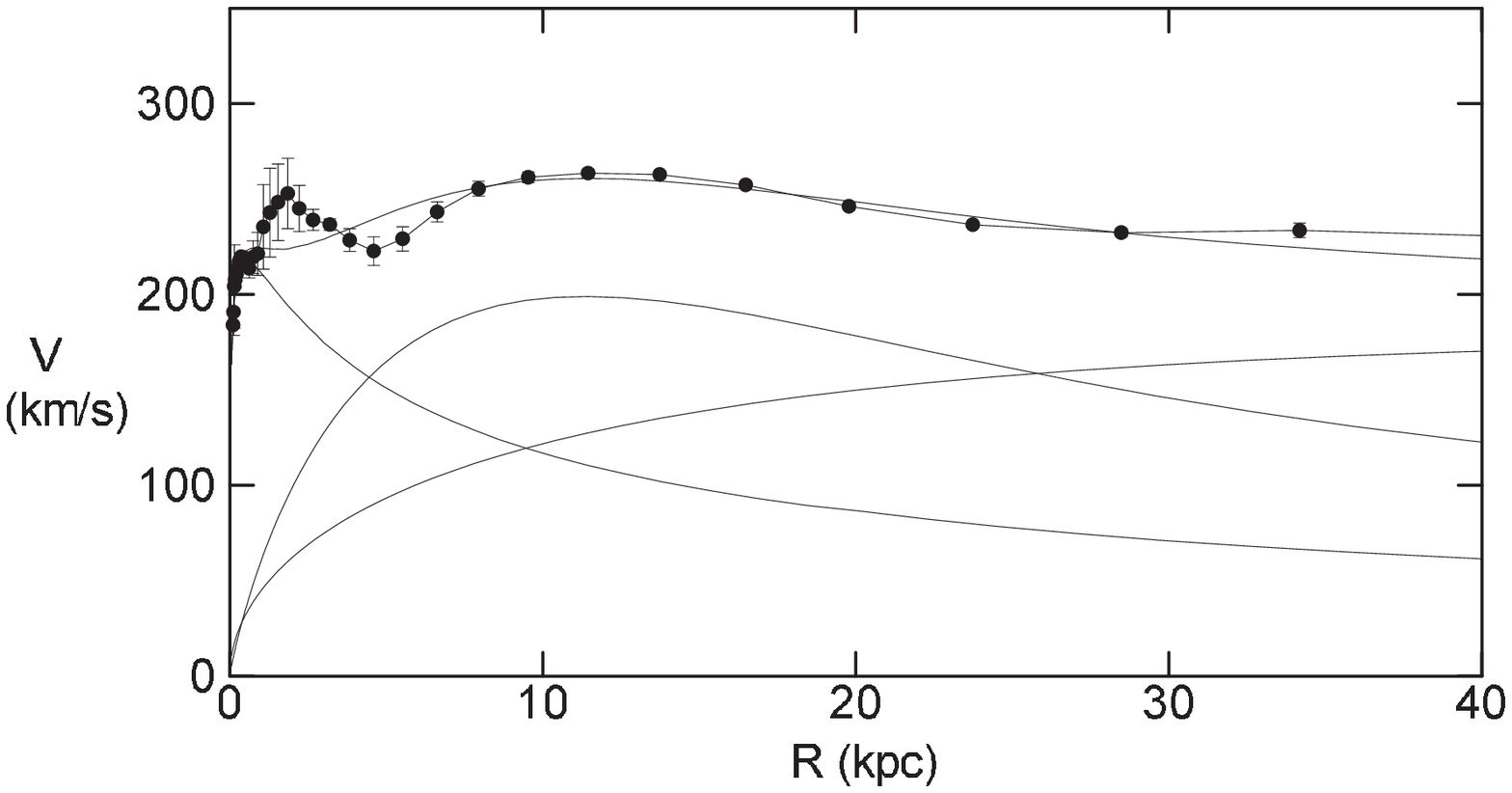}  
(b)\includegraphics[width=8cm]{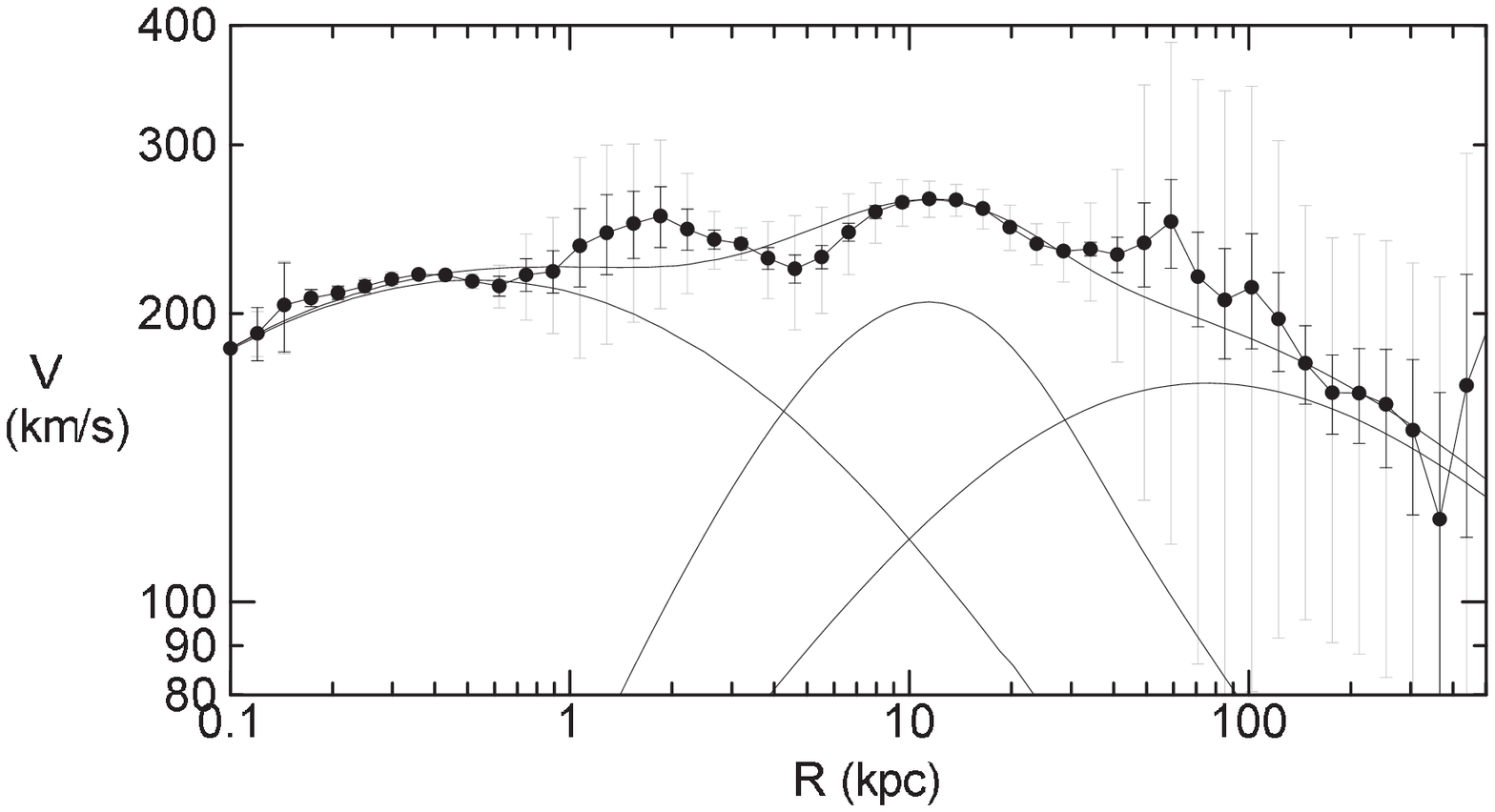}    
\end{center}
\caption{Least-$\chi^2$ fit of the GRC of M31 by the bulge, disk and dark halo components in (a) linear and (b) logarithmic scalings.  } 
\label{fit385}   

\begin{center}    
(a)\includegraphics[width=8cm]{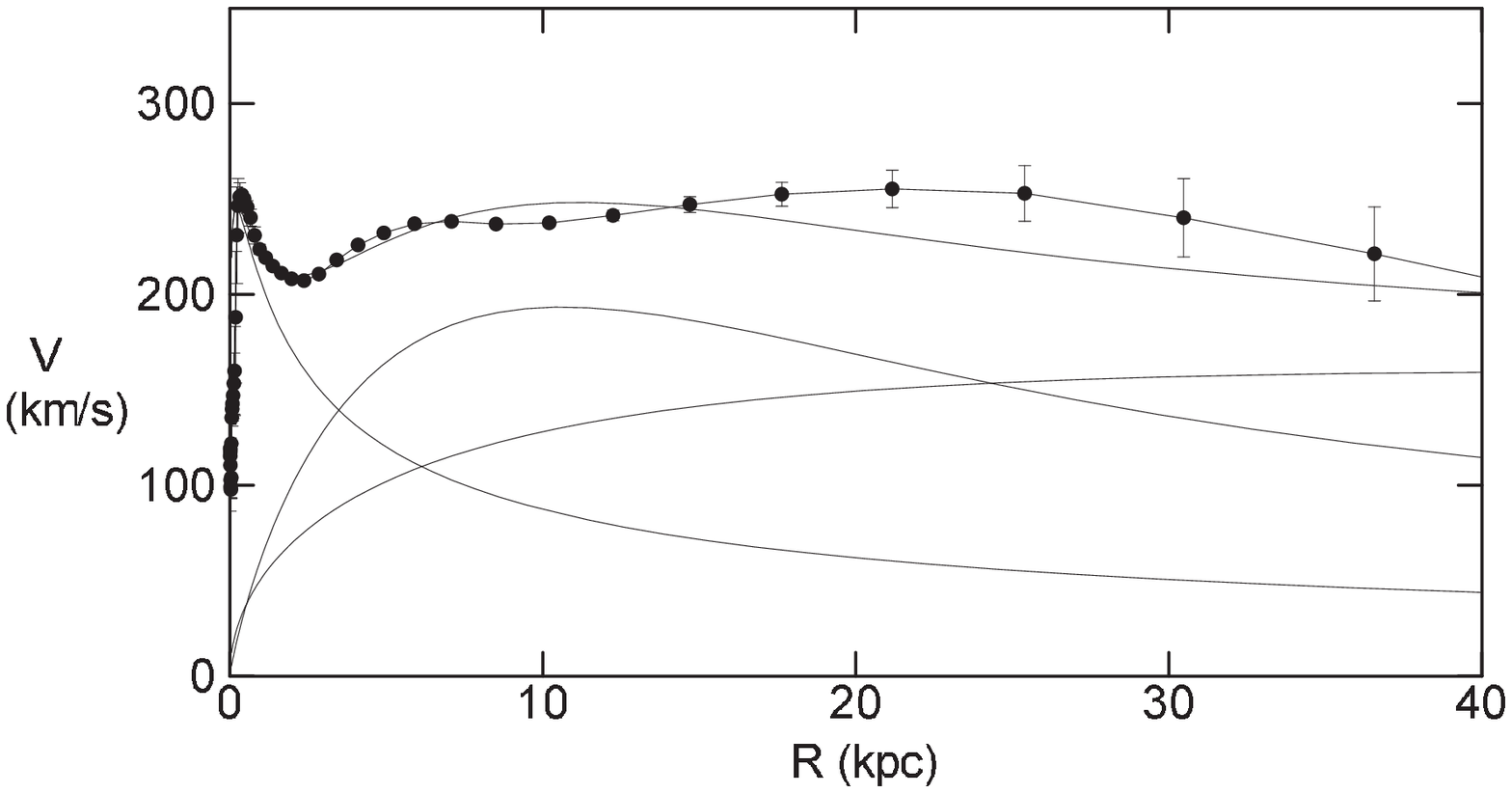}  
(b)\includegraphics[width=8cm]{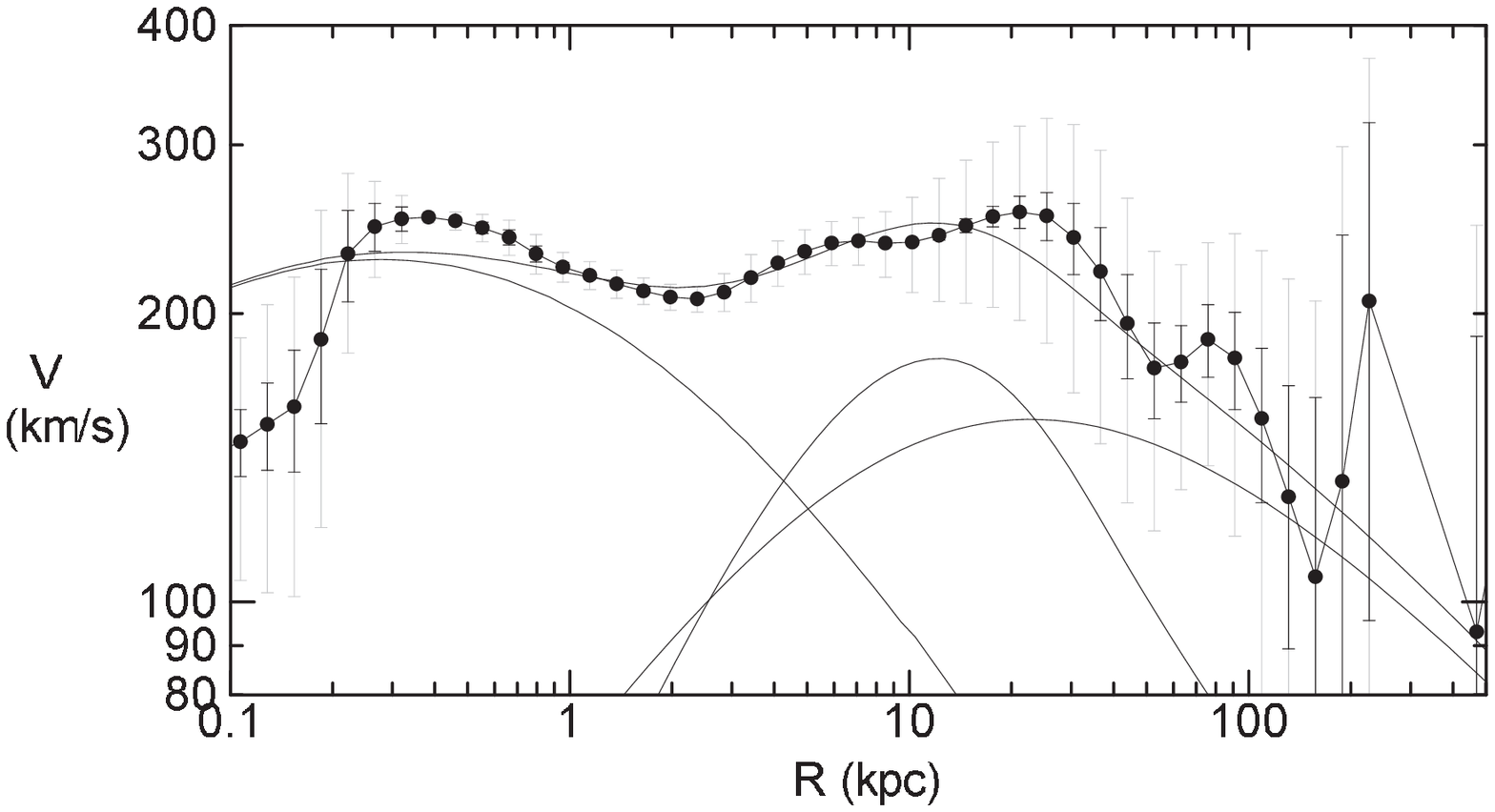}    
\end{center}
\caption{Same as figure \ref{fit385}, but for the revised new GRC of Milky Way using the same procedure as for M31. } 
\label{fit385mw}  
\end{figure}

Applying the fitting method described in Sofue (2012) to the GRC of M31 and the Galaxy, we searched for the best-fit parameters of the bulge, disk and dark halo. The free parameters are $\Mb,~ \ab,~ \Md,~ \ad,~ \rho_0$ and $h$. Fitting radii were taken to be $R_1=0$ to $R_2=$ 20 kpc for the bulge, 0 to 40 kpc for disk, and 1 to 385 kpc for the dark halo. The outer boundary for the halo fitting corresponds to the half distance between the two galaxies.  Figures \ref{fit385} and \ref{fit385mw} show the thus obtained fitting results compared with the used GRCs.  Table 2 shows the best-fit parameters for individual mass components. 

\begin{figure}
\begin{center}  
(a)\includegraphics[width=8cm]{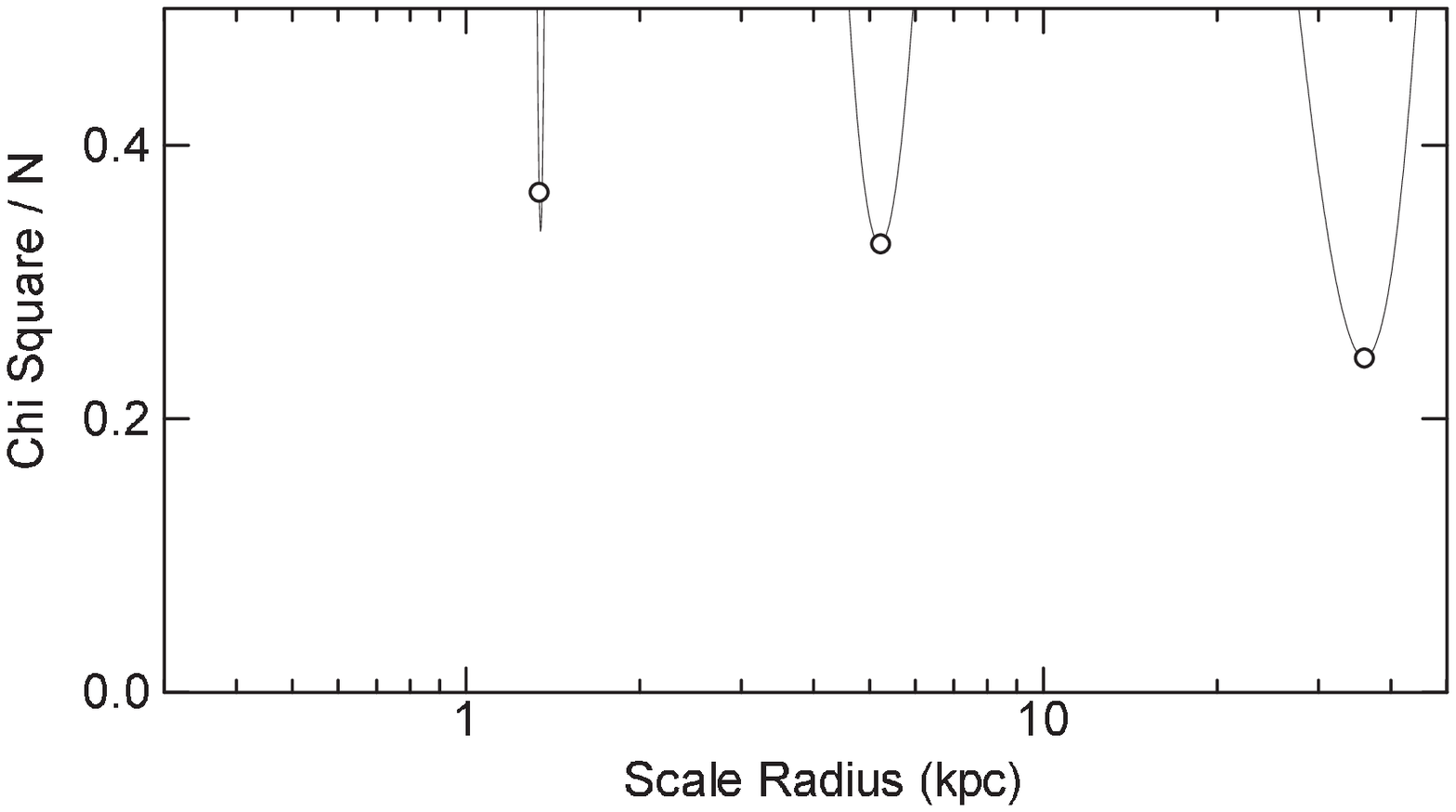}   
(b)\includegraphics[width=8cm]{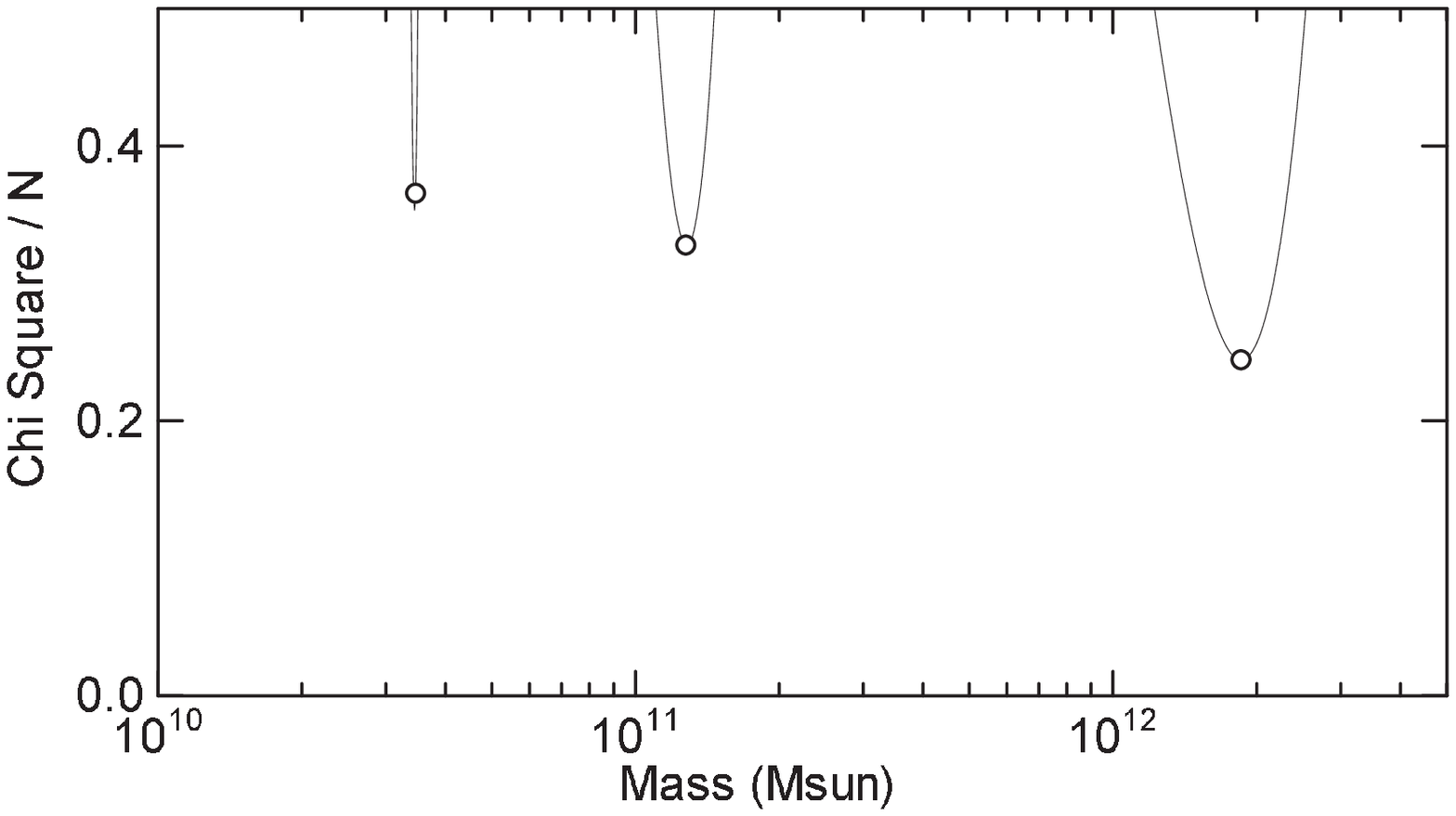}   
\end{center}
\caption{(a) Values of $\chi^2/N$ for M31 as functions of $\ab$, $\ad$, and $h$ around the least values marked by circles. From left to right: bulge, disk, and dark halo, respectively. (b) Same, but for the masses $\Mb$, $\Md$ and $\Mh$$_{:385}$.  }  
\label{XiA}  

\begin{center}  
(a)\includegraphics[width=8cm]{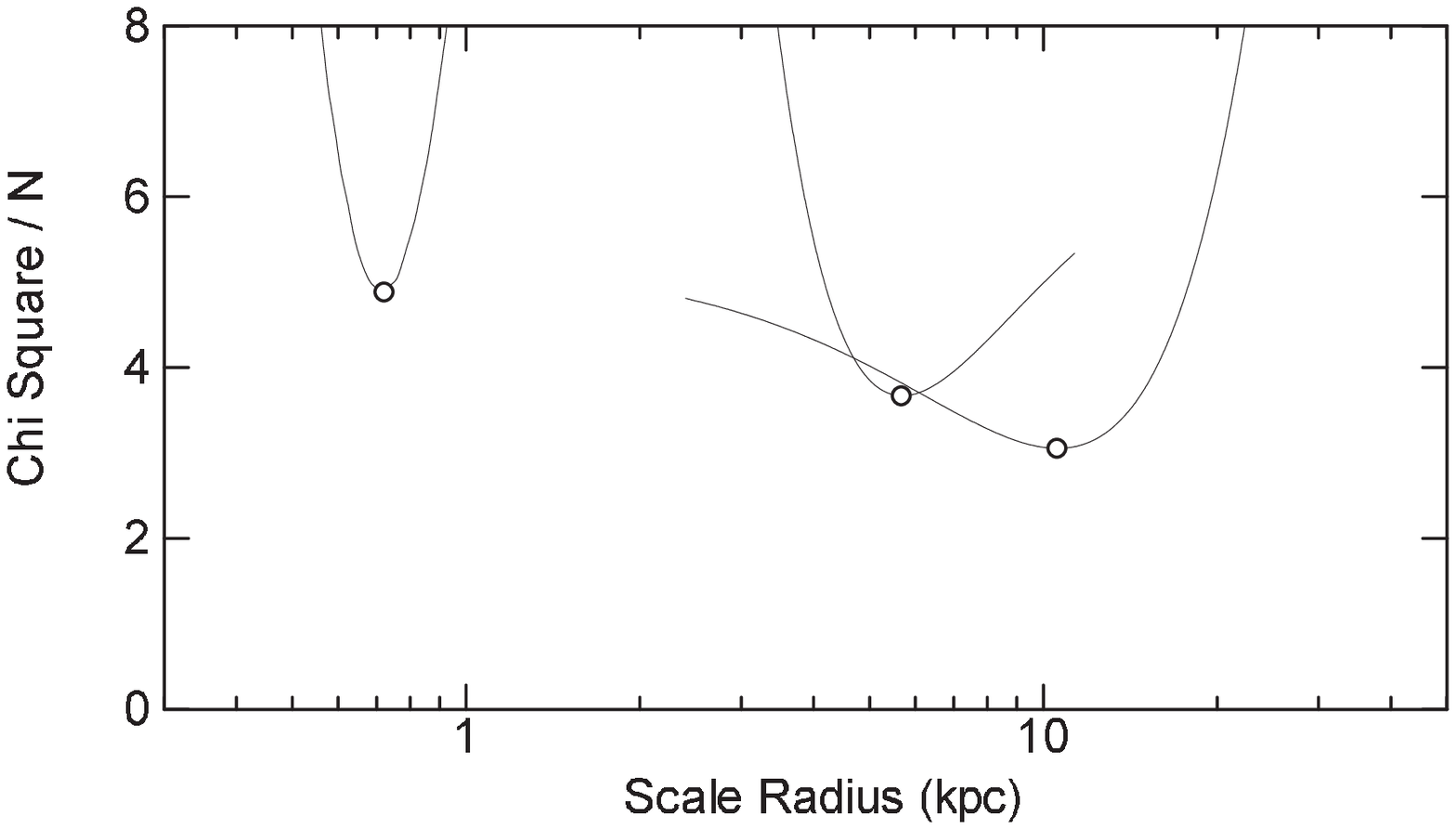}   
(b)\includegraphics[width=8cm]{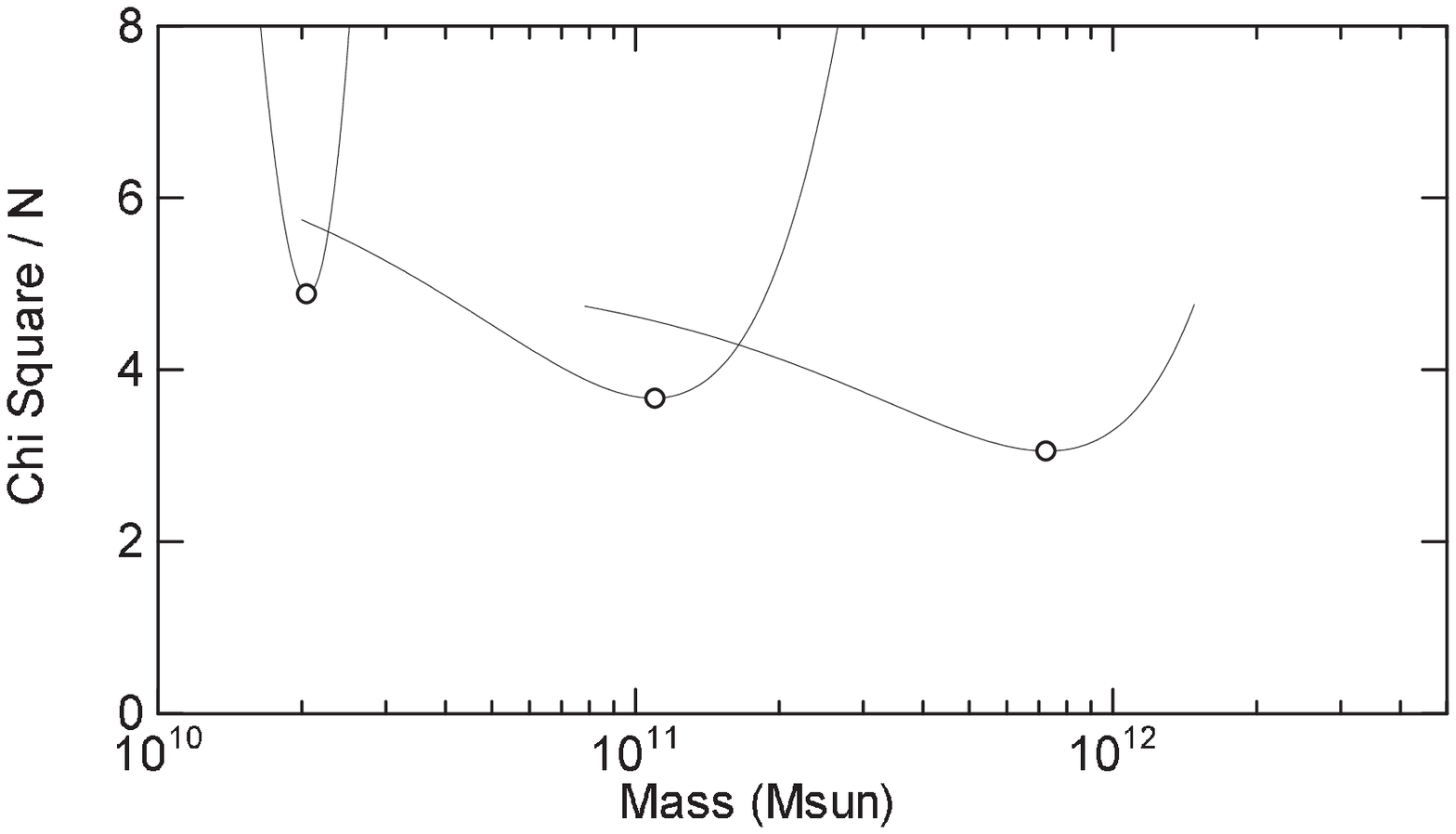}   
\end{center}
\caption{Same as figure \ref{XiA}, but for the Milky Way using the newly determined revised GRC. }  
\label{XiAmw} 
\end{figure}

Figures \ref{XiA} and \ref{XiAmw} show the behaviors of $\chi^2/N$ around the least values, where $\chi$ is defined by $\chi^2=\Sigma [V_{\rm c}(R_i)^2-V_{\rm o}(R_i)^2]/s_d^2$ with o and c abbreviating observed and calculated values, respectively, and $N$ is the number of fitting points.  For the dark halo we present the total dark mass, $\Mh$ $_{:385}$, corresponding to $h$ and $\rho_0$. Since the fitting areas and $N$ for bulge, disk, and halo are different, the minimum $\chi$ values are different among the components. The error of each fitted parameter was evaluated as the range that allows for an increase of the $\chi^2$ value by 10\% above the least value.

\def\m11{10^{11}\Msun}
\begin{table*} 
\bc
\caption{The best fit dynamical parameters for M31 and the Galaxy$\dagger$} 
\begin{tabular}{llll}
\hline\hline 
Component&Parameter&M31 &Milky Way \\
\hline  
\hline  

 Bulge & $a_{\rm b}$ (kpc)& $ 1.35\pm  0.02$ & $  0.87\pm  0.07$ \\
 & $M_{\rm b} (\m11)$ & $  0.35\pm  0.004$ & $  0.25\pm  0.02$ \\
\hline
 Disk & $a_{\rm d}$ (kpc) & $  5.28\pm  0.25$ & $  5.73\pm  1.23$ \\
 & $M_{\rm d} (\m11)$ & $  1.26\pm  0.08$ & $  1.12\pm  0.40$ \\
\hline
 NFW Halo& $h$ (kpc) & $ 34.6\pm  2.1$ & $ 10.7\pm  2.9$ \\
 & $\rho_0 (10^{-3}\Msun{\rm pc}^{-3})$ & $ 2.23\pm 0.24$ & $ 18.2\pm 7.4$ \\
 & $\rho_{\rm 8~kpc} (10^{-3}\Msun {\rm pc}^{-3})$ & $  6.36\pm  0.70$ & $  7.93\pm  3.24$ \\
 & = (in energy density: GeV ${\rm cm}^{-3}) $ & $  0.24\pm  0.03$ & $  0.30\pm  0.12$ \\
 & $M_{\rm h: 200}(\m11)$ & $ 12.3\pm  2.6$ & $  5.7\pm  5.1$ \\
 & $M_{\rm h: 385}(\m11)$ & $ 18.3\pm  3.9$ & $  7.3\pm  6.7$ \\
\hline
 Total Mass& $M_{\rm tot: 200}(\m11)$&$ 13.9\pm  2.6$ & $  7.0\pm  5.1$ \\
 & $M_{\rm tot: 385}(\m11)$&$ 19.9\pm  3.9$ & $  8.7\pm  5.1$ \\
\hline
  Bulge&$\chi^2_{\rm b}/N$ ($R_1-R_2$ kpc)&0.36 (0.0-20.0)&3.5 (0.0-20.0)\\
  Disk&$\chi^2_{\rm d}/N$ ($R_1-R_2 $ kpc)&0.33 (0.0-40.0)&3.6 (0.0-40.0)\\
  Halo&$\chi^2_{\rm h}/N$ ($R_1-R_2 $ kpc)&0.25 (0.0-385.0)&3.0 (0.0-385.0)\\

\hline
\end{tabular}
\ec
$\dagger~ \Mh$$_{:200}$, $\Mh$$_{:385}$,$M_{\rm tot: 200}$ and $M_{\rm tot: 385}$  are dark halo and total masses within $R=200$ and 385 kpc, respectively;  
$\rho_{8{\rm kpc}}$ is a local value at $R=8$ kpc both in mass and energy densities;
$R_1$ and $R_2$ are start and end radii for fitting.
\label{tabfit} 
\end{table*} 


\section{Discussion} 

\subsection{Summary}
We constructed GRCs of the Andromeda galaxy M31 and the Milky Way Galaxy for wide regions from the centers to the dark halo edges (figures \ref{rc385} and \ref{rc385mw}).  The GRC for the Milky Way was revised by adopting the most recent Solar rotation velocity and applying the same method as for M31. As stressed below, the dark halos of both galaxies are well represented by the NFW density profiles.

\begin{figure}
\begin{center} 
\includegraphics[width=8cm]{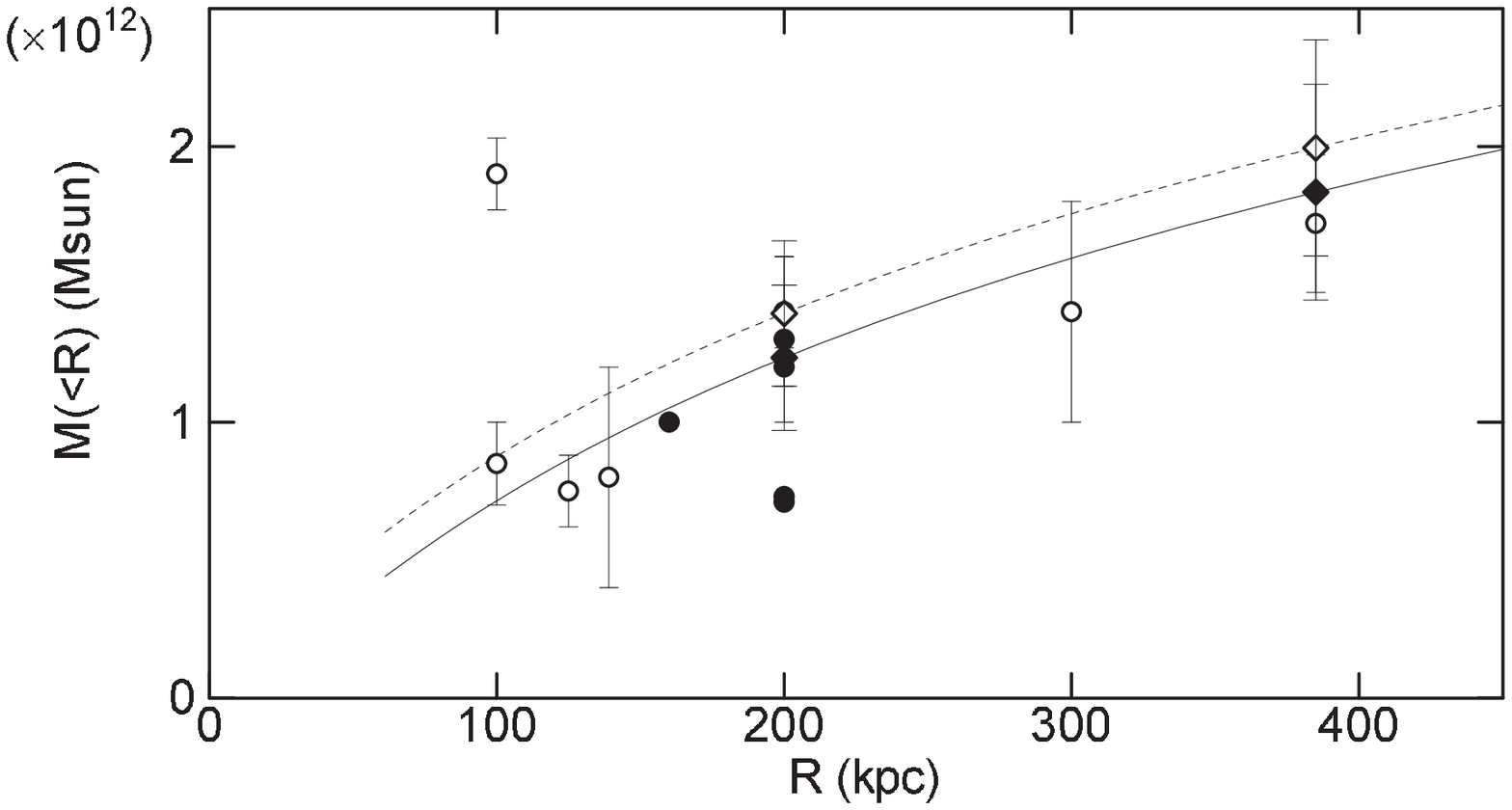} 
\end{center}
\caption{Enclosed dark (filled) and total masses (open circles) within $R$ of M31 obtained by the authors listed in table 3.    Upper (dashed) and lower (fill) lines show $M_{\rm tot}(R)$ and $\Mh(R)$, respectively, calculated using the fitted parameters of M31. Big diamonds on the calculated lines indicate the present result.} 
\label{figcomp} 
\end{figure}

By the least-$\chi^2$ fitting to the obtained GRCs up to radius of 385 kpc (figures \ref{fit385} and \ref{fit385mw}), we determined the galactic parameters for the bulge, disk, and dark halo, as listed in table 2. Our result for M31 is consistent with those obtained by the other authors in the decade, as compared in table 3 and figure \ref{figcomp}.

\subsection{The reality of NFW model}
 The fitting result in figures \ref{fit385} and \ref{fit385mw} proves that the NFW profile (Navarro et al. 1995) can be a realistic approximation to represent observed dark halos. We emphasize that the analysis of GRCs covering regions as wide as several hundred kpc around the galaxies is essential to discriminate the right model among the three types of dark halo models: the isothermal model predicting flat rotation in the outermost region as $V_{\rm rot}\propto {\rm const}$, the NFW model predicting slowly decreasing rotation as $\propto({\ln}~R/R)^{-1/2}$, and Plummer- or exponential-type models predicting Keplerian decrease as $\propto R^{-1/2}$ at large radii.

\begin{table} 
\caption{Comparison of derived masses of dark halo and total mass of M31 with other works in the decade.} 
\bc
\begin{tabular}{lll} 
\hline
\hline
Authors &  Mass&  in $ 10^{11}\Msun$   \\
\hline 
Evans et al. (2003) & $M_{{\rm tot}:100}$& 7-10  \\ 
Ibata et al. (2004)& $M_{{\rm tot}:125}$ & $7.5\pm 1.3 $ \\
Geehan et al. (2006) &$\Mh$$_{:200}\dagger$& 7.1 \\
Seigar et al. (2008) &$\Mh$$_{:200}\dagger$& 7.3  \\
Lee et al. (2008) & $M_{{\rm tot}:100}$ & $19.0\pm 1.3$ \\ 
Chemin et al. (2009) &$\Mh$$_{:160}$& 10.0  \\
Corbelli et al. (2010) &$\Mh$$_{:200}\dagger$& $13\pm 3$ \\ 
Watkins et al. (2010) & $M_{{\rm tot}:300}$& $14\pm 4$ \\
van der Marel et al. ('12) & $M_{\rm tot: Vir(385)}$& $17.2 \pm 2.5$ \\
Tollerud et al. (2012) & $M_{{\rm tot}:139}$ & $8 \pm 4$ \\ 
Tamm et al. (2012) & $\Mh$$_{:200}\dagger$& 11.3--12.7  \\ 
Fardal et al. (2013) & $M_{{\rm tot}:200}$ & $19\pm 5$ \\
Veljanovski et al. (2014) &$\Mh$$_{:200}\dagger$& 12--16  \\
\hline
This work (2015) & $\Mh$$_{:200}\dagger$& $12.3 \pm 2.6$\\
--- & $\Mh$$_{:385}$& $18.3 \pm 3.9$  \\
--- & $M_{{\rm tot}:200}$& $13.9 \pm 2.6$ \\
--- & $M_{{\rm tot}:385}$& $19.9 \pm 3.9$  \\ 
\hline
Average $<\Mh$$_{:200}>$$^\dagger$ &&$10.8\pm 3.0$\\
Average $<M_{{\rm tot}{:100-385}}>$ & &$13.7\pm 5.2$\\
\hline

\end{tabular}
\ec 
\label{tabcomp} 
\end{table}

\subsection{Similarity of the GRCs in M31 and the Milky Way and difference in the dark halo masses}

In figure \ref{M31vsMW} the GRC of M31 is compared with the revised GRC of the Milky Way, and the fitting results are compared in table 2. There is remarkable similarity of rotation curves inside the disk regions as well as in the dark halos in their shapes and amplitudes. However, the fitted dark halo mass of M31 is about twice that of the Galaxy and the parameters are also different, although the bulge and disk parameters are similar between the two galaxies. 

\subsection{Revised GRC of the Milky Way}
The fitting accuracy in the Milky Way is not satisfactory, particularly for the bulge component (figure \ref{XiAmw}), which is mainly due to the fact that the Galactic bulge is composed of multiple components, not well represented by the de Vaucouleurs law (Sofue 2013). However, we used the common model to M31 in order to compare the global structures of the two galaxies. Hence, the innermost structure of the Milky Way may not be taken serious here, while it does not affect the result for the disk and dark halo.

The here determined scale radii and masses for the Milky Way are systematically greater than those obtained in our previous work (Sofue 2013). This is because of the adopted larger rotation velocity of the Sun, which causes significant increase of the disk rotation velocities, as well as to the revised correction factor of radial velocities of non-coplanar objects.

\subsection{Baryonic fraction}
The ratio of the total mass of the bulge and disk to the dark matter mass, 
$\Gamma=(M_{\rm b}+M_{\rm d})/\Mh$$_{:385}$, is a measure of the baryonic to dark mass ratio. The values in table 2 yield the ratio of $\Gamma\sim 0.087$ for M31, and 0.25 for the Galaxy. These values may be compared with the cosmological value of 0.19 (=4.5\%/24\%) from the WMAP observations (Spergel et al. 2003). This implies that M31 is dark mater dominant compared to the cosmological value, whereas the Galaxy is baryon exceeded.

\subsection{Relation to the Local Group}
The systemic velocity of M31 with respect to the Milky Way of $\sim 100$ \kms (Courteau and van den Bergh 1999) can be translated to a statistically possible mutual velocity by multiplying $\sqrt{3}$ to yield $\sim 170$ \kms, if we assume that the two galaxies' motions are random. In order for the two galaxies to be bound, a reduced total mass of the whole system is required to be greater than $\sim 5\times 10^{12} \Msun$. Since both M31 and the Galaxy are far less massive, $\sim 3\times 10^{12}\Msun$ in total, than this binding mass, there remain two possibilities either that the Local Group contains dark mass as massive as $\sim 2 \times 10^{12}\Msun$, or that the two galaxy systems are not gravitaionally bound. 

In the former case, our view that the Local Group can be recognized as one system may not be changed. However, the dark mass must be concentrated around the center of mass of the Local Group, so that the M31 and the Galaxy groups, both associated with own dark halos and embedded satellite galaxies and globular clusters, are not tidally disrupted.

In the latter idea, we need no further assumption of dark mass in the Local Group, but it is required that M31 and the Galaxy and their satellite galaxies have been two individual bound systems since the formation. In fact, Sawa and Fujimoto (2005) proposed a model that dwarf galaxies of the Local Group are bound either to M31 or the Milky Way, composing two individual bound groups, and the two groups are orbiting around each other and tidally interacting.


\begin{appendix}
\section{Pseudo rotation velocity in pressure-supported rotating system}

\def\vector#1{\mbox{\boldmath $#1$}}
\def\vv{\vector{v}}
\def\vV{\vector{V}}
 
If there exists systematic rotation $\vV_{\rm rot}$ in the halo, the   velocity $V=|\vector{V}|$ of a satellite galaxy and/or globular cluster (hereafter, particle)  is expressed as  
\begin{equation}
V^2=(\vV_{\rm rot}+\vV_{\rm ran})^2
=V_{\rm rot}^2+2 \vV_{\rm rot} \cdot \vV_{\rm ran}+V_{\rm ran}^2,
\end{equation}
where $\vV_{\rm ran}$ is the velocity corresponding to pressure-support term.
When $V^2$ is averaged around the galaxy, the crossing term disappears because of randomness of $\vV_{\rm ran}$ and axisymmetry of $\vV_{\rm rot}$, yielding 
\begin{equation}
<V^2>=<{V}_{\rm rot}^2>+<{V}_{\rm ran}^2>.
\end{equation}
The first term is related to the apparent rotation velocity projected on the sky, $v_{z:{\rm rot}}$ as
\be
v_{z:{\rm rot}}= V_{\rm rot}{\rm sin}~i~ {\rm cos}~\theta.
\ee
Here, $i$ is the inclination angle of the rotation axis, and $\theta$ is the azimuthal angle of the particle from the major axis. Knowing that $i$ and $\theta$ are independent and $ <{\rm cos^2}\theta>=1/3$ by averaging over $\theta$ from 0 to $\pi/2$, we have
\be
< V^2_{\rm rot}> = {3 \over {\rm sin}^2 i} < v^2_{z:{\rm rot}}>.
\ee
 Replacing
$<{V}_{\rm ran}^2>$ by $3<v_{z:{\rm ran}}^2>$, where $v_{z:{\rm ran}}$ is the $z$-directional component of $\vV_{\rm ran}$, we obtain
\be
<V^2>=3<v^2_{z:{\rm ran}}>+{3 \over {\rm sin}^2 i}<v^2_{z:{\rm rot}}>.
\ee
Writing the observable $z$-directinal velocity as $v_z=v_{z:{\rm ran}}\pm v_{z:{\rm rot}}$, and remembering that the crossing term $<\pm 2v_{z:{\rm ran}} v_{z:{\rm rot}}>$ reduces to zero for randomness of sign of $v_{z:{\rm ran}}$, we obtain
\be
<v^2_z>=<v^2_{z:{\rm ran}}>+< v^2_{z:{\rm rot}}>.
\ee  
 Hence, we have
 \be
<V^2>=3<v^2_z>+3{\rm cot}^2 i<v^2_{z:{\rm rot}}>.
\label{Asini}  
\ee 
For an edge-on case, $i=90^\circ$, we have 
\be
<V^2>=3<v^2_z>,
\ee
the same as for random motion (Limber and Mathews 1960). 
If we assume the same inclination as the main disk of M31, $ i=77^\circ$ to $78^\circ$,we have
\be
<V^2>=3<v^2_z>+ \alpha <v^2_{z:{\rm rot}}>
\ee
with $\alpha \simeq 0.15$.  Since  $<v^2_{z:{\rm rot}}>$ is smaller than $<v^2_z>$, we may neglect the second term, obtaining
\be
<V^2> \simeq 3<v^2_z>. 
\ee
Finally, we define the pseudo rotation velocity as
\be
V=\sqrt{<V^2>}=\sqrt{3<v^2_z>}.
\ee

\end{appendix}

\end{document}